\documentclass[preprint,aps,amsmath]{revtex4}
\usepackage{graphicx}

\begin{document}

\draft

%\wideabs{ % Remove wideabs before submittal
\title{Comparison of the roughness scaling 
				of the surface topography of Earth and Venus.}

\author{G. E. Crooks}
\address{Dept. of Chem., University of California Berkeley, Berkeley, CA}
\author{Y. Bar-Yam}
\address{New England Complex Systems Institute, Cambridge, MA }
\author{S. V. Buldyrev, H. E. Stanley}
\address{Polymer Center, Dept. of Physics, Boston University, Boston, MA}

\date{Nov. 2, 1998}

\begin{abstract}
	We report the scaling behavior of the Earth and Venus over a wider 
	range of length scales than reported by previous researchers.  All  
	landscapes (not only mountains) together follow a consistent 
	scaling behavior, demonstrating a crossover between highly 
	correlated (smooth) behavior at short length scales (with 
	a scaling exponent 
	$\alpha$=1) and self-affine behavior at long length scales 
	($\alpha$=0.4).  The self-affine behavior at long scales is 
	achieved on Earth above 10 km and on Venus above 50 km.
\end{abstract}

\pacs{91.10.Jf, 96.35.Se, 96.35.Gt, 93.30.Hf}
%}
\maketitle

It has been suggested based on a number of studies that horizontal 
transects of mountain ranges follow self-affine 
scaling\cite{Briggs89,Gilbert89,Malinverno89,Matsushita89a,%
Matsushita89b,Huang89,Huang90,Kucinskas92,Chase92,Dietler92,%
Ouchi92,Klinkenberg92,Cox93,Rigon94,Czirok94,Turcotte97}. 
 However, quite different values of the scaling exponent have been 
proposed.  Typically, it has been found that scaling exponents at 
shorter length scales are higher than scaling exponents at longer 
length scales.  A consistent picture of the scaling behavior of the 
Earth's topography accounting for these diverse results has not yet 
been provided.  We find, by studying a larger range of length scales, 
that all landscapes (not only mountains) together follow a consistent 
scaling behavior, demonstrating a smooth crossover between a 
relatively smooth, correlated landscape at short length scales (with a 
scaling exponent $\alpha=1$) and self-affine behavior at long length 
scales ($\alpha=0.4$).  Contrary to expectations that mountain ranges 
are self-affine over all scales, we find that only above distances of 
10 km does a self-affine scaling with a unique exponent $\alpha=0.4$ 
apply.  The topography of Venus shows a similar crossover at a larger 
length scale of 50 km to self-affine scaling that extends to $7\times 
10^{3}$ km.

To investigate the scaling of topography, it is necessary to analyze 
the Earth's surface over a wide range of lengths.  Topographic data 
for the US have been collected by the US Geological Survey\cite{USGS}.  
While data are available for other regions of the Earth, this is the 
highest resolution data over the widest area publicly available.  
Elevations for the contiguous 48 States are provided in 1 degree 
square blocks (approximately 100 km by 100 km), at a horizontal 
resolution of 3 arc-seconds ($\approx$100 m).  This provides about 
$10^{9}$ data points covering an area of $7.5\times10^{6}{\rm 
km}^{2}$.

For a self-affine landscape the standard deviation of surface 
elevation $W$ should scale with sample size (linear dimension) $L$ as

\begin{equation}
 	W \sim L^{\alpha} ,
\end{equation}
\noindent where $\alpha$ is the roughness scaling exponent.  A number 
of methods have been used to obtain the scaling behavior.  We 
calculate $W$ for square areas of increasing edge length $L$, and then 
average over the entire region.  Averages of the scaling behavior of 
linear transects taken through the same dataset provided essentially 
similar scaling behavior.  We use real space rather then Fourier space 
evaluation of the standard deviation since Fourier transforms impose 
periodic boundary conditions and thus discontinuities which result in 
distortions of the scaling behavior at short length 
scales\cite{Austin94}.

Figure 1 shows the calculated values of $W(L)$ for a random selection of 
\begin{figure}
	\includegraphics[width=12 cm]{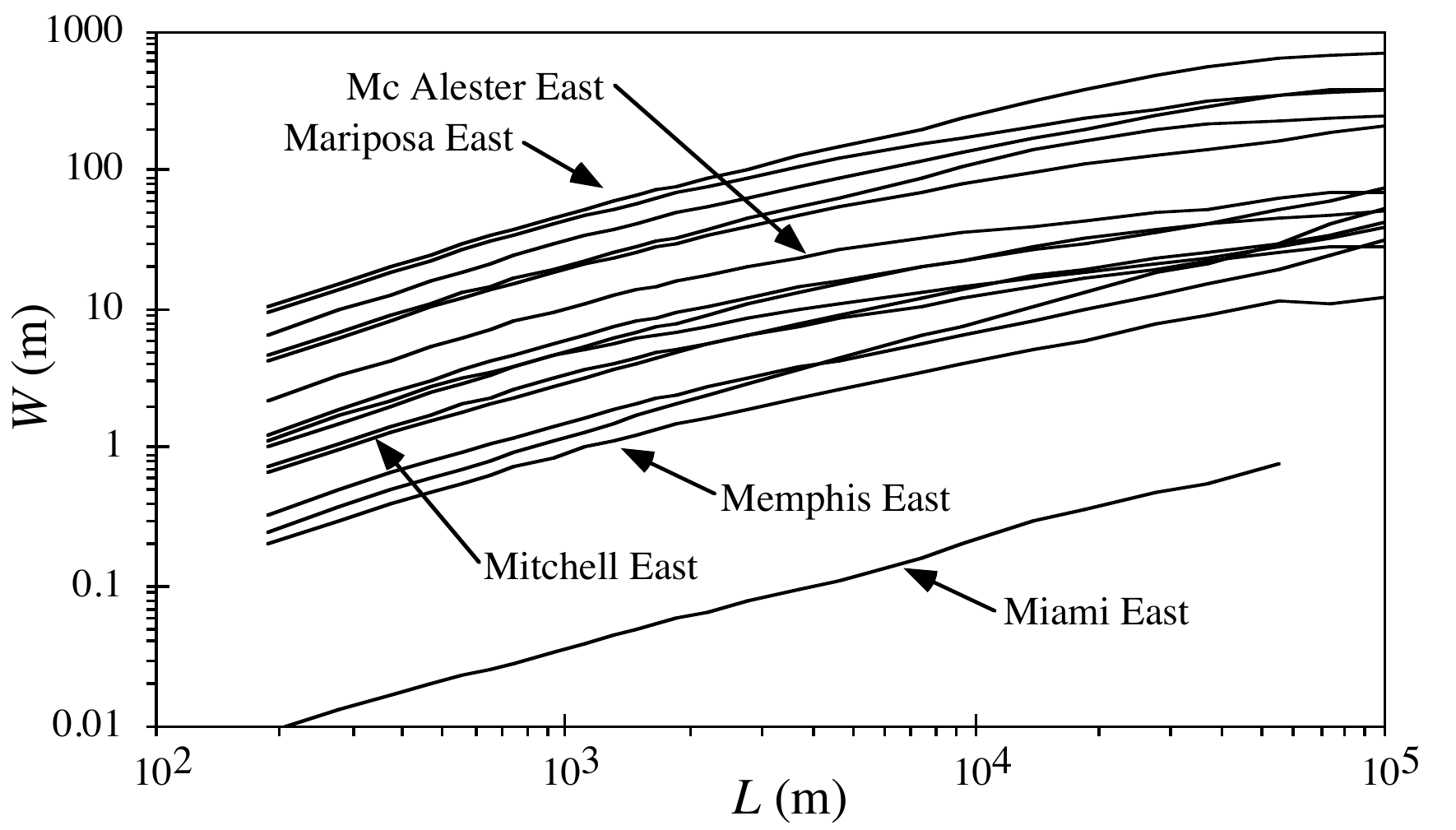}
	\caption{Standard deviation $W$ of elevations in randomly selected 
	100 km by 100 km regions of the US. Several locations with very 
	different topographies, have been labeled.  Mariposa East covers 
	part of the Sierra Nevada mountains, and contains Kings Canyon 
	national park.  McAlester East is in eastern Oklahoma.  Mitchell 
	East is in southeast South Dakota.  Memphis East covers the 
	southwest corner of Tennessee, and is bisected by the Mississippi 
	river.  Miami East covers part of the Florida everglades.  We 
	calculate $W$ for square areas of increasing edge length $L$, and then 
	average over the entire region.  The curves are nearly parallel, 
	illustrating the remarkable universality of roughness scaling 
	despite the wide variation in absolute roughness.  The larger 
	variation at larger values of $L$ can be attributed to averaging 
	over fewer samples at these lengths.  }
	\label{Fig:USA}
\end{figure}
1 degree by 1 degree regions.  The curves represent a great range of 
landscapes.  The top curve, Mariposa East (37-38 N, 118-119 W), covers 
part of the Sierra Nevada range including Kings Canyon National Park 
in California.  This extremely mountainous terrain has a high point at 
4341 m above sea level.  The bottom curve in Fig.~1 represents data 
from the Everglades, an extremely flat terrain (Miami East, 25-26 N, 
80-81 W).  Intermediate curves are indicated in the figure.  This 
selection of landscapes covers 3 orders of magnitude in roughness.

Remarkably, it can be seen that most of the landscapes have similar 
scaling behavior.  In particular, scaling properties are not limited 
to mountainous regions, but apply to plains and flatlands as well.  
Scaling exponents calculated for each region from $L$=40 km to 120 km 
are broadly distributed with a mean of 0.42, and a standard deviation 
of 0.2.  There appears to be no significant correlation ($r=0.17$) 
between $\alpha$ and the roughness of the landscape (determined by the 
value of $W$ at $L=120$ km).  There is a weak correlation between 
geographically neighboring regions.

Figure 2 shows results for the entire continental US. There is a 
\begin{figure}
	\includegraphics[width=12 cm]{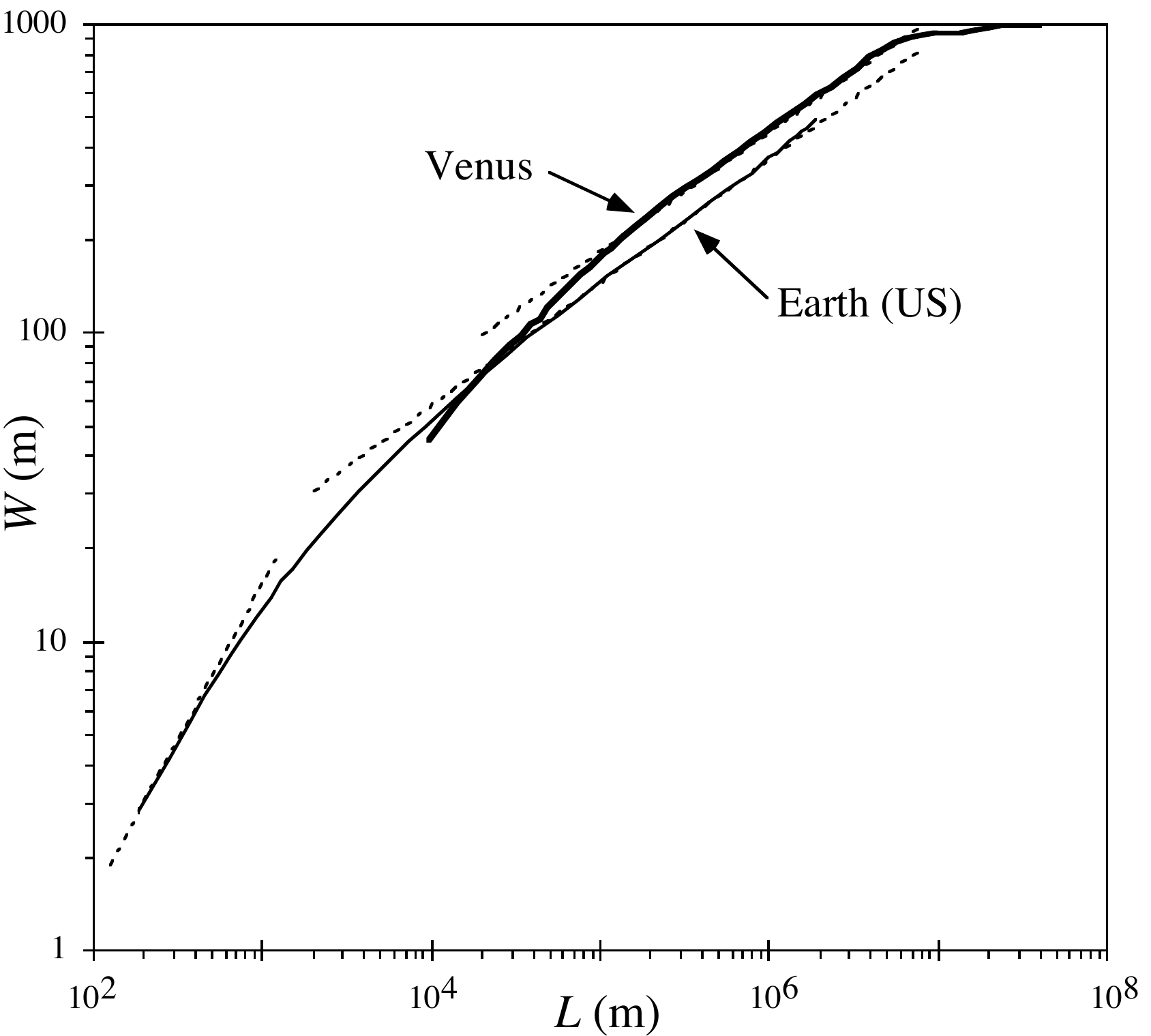}
	\caption{Standard deviation of elevations of the entire 
	continental US (thin solid line) and of the surface of Venus 
	(thick solid line).  The US results are calculated using square 
	areas as in Fig.~1.  Due to surface curvature of Venus at longer 
	length scales, we calculate the Venus results for circular areas 
	of diameter $L$. Square areas produce similar results.  For the US 
	data, we find a crossover between highly correlated smooth behavior 
	($\alpha$=1) at the 
	shortest length scales, and self-affine behavior at the longest 
	scales.  The upper dashed line is fit to the data between 100 and 
	1000 km and has an exponent $\alpha$=0.40.  The lower dashed line has an 
	exponent of $\alpha$=1.0.  The Venus data have a similar crossover.  The 
	corresponding dashed line is fitted between 200 to 2000 km and has 
	an exponent of $\alpha$=0.39.  }
	\label{Fig:EV}
\end{figure}
variation of $\alpha$ with length scale consistent with previous studies.  
The asymptotic scaling regime is reached only at approximately 5 km 
and the scaling exponent (Fig.~3) has a value of 0.4.  This exponent 
corresponds to the value predicted for scaling of self-affine surfaces 
that are growth fronts\cite{Kardar86}. It is not obvious, however, that 
these models should apply to the Earth's topography.  Nevertheless, scaling 
arguments suggest that essentially all dynamic surfaces with up-down 
asymmetry are part of a universality class with this scaling exponent. 
Correlations exist up to the maximum length scale of the data.  
However, at shorter length scales the landscape is smooth and approaches 
a scaling exponent of $\alpha$=1. This scaling exponent is consistent 
with a highly correlated 
affine surface or even a linear surface on short length scales. 
It reflects the smoothness of both mountain slopes and other terrain 
at these length scales.

Our results are consistent with the overall pattern of observations by 
other researchers.  For example, Dietler and Zhang\cite{Dietler92} 
performed an analysis for Switzerland, an area of $7\times10^{4} 
{\rm{km}}^{2}$ with a lower resolution of 250 m, and by fitting a 
single line to values below 5 km they obtained $\alpha\approx 0.57$.  
They also indicate that higher resolution data led to still larger 
slopes.  Above this scale their data suggested $\alpha=0.27$, whose 
low value may be attributed to the small number of independent 
samples, consistent with the range of values found in the data we 
studied.  Turcotte's analysis\cite{Turcotte87,Turcotte97} of the large 
scale Earth scaling behavior yielded a value of $\alpha$=0.5.  His 
analysis combines bathymetric and topographic data.  We have 
separately analyzed bathymetric data, which dominates the large scale 
Earth scaling, and find its large length scaling exponent to be close 
to 0.5 in agreement with his results.  Our analysis, however, 
indicates a real difference between bathymetric and topographic large 
scale scaling exponents.  The bathymetric data is also rougher on an 
absolute scale.  When considering the entire earth, the inclusion of 
the continental shelf in Turcotte's analysis makes his absolute 
roughness still greater.  Our results are also consistent with 
Turcotte's analysis\cite{Huang90} of the scaling behavior of the state 
of Oregon, for which he reports $\alpha=0.414$, and the state of 
Arizona for which he reports $\alpha=0.41$.

In order to explore the universality of these results, we also study 
the topography of Venus using Magellan data.\cite{Towheed93} Venus 
data does not suffer from a need to consider oceans.  We find that the 
landscape of Venus is self-affine with $\alpha$=0.4 for length scales 
between 50 km and $7\times 10^{3}$ km (Fig.~2).  Above $7\times 
10^{3}$ km there are inherent correlations due to the maximum mountain 
height, determined by the gravitational force and the strength of the 
substrate.  For length scales less than 50 km we find a larger scaling 
exponent as for the Earth data.  We note that care must be taken in 
analysis of the Venus data because of the existence of deep gorges.  
Because of these gorges the higher moments of the surface topography 
have lower scaling exponents; however, the gorges do not contribute 
substantially to the curves in Fig.~3.
\begin{figure}
	\includegraphics[width=12 cm]{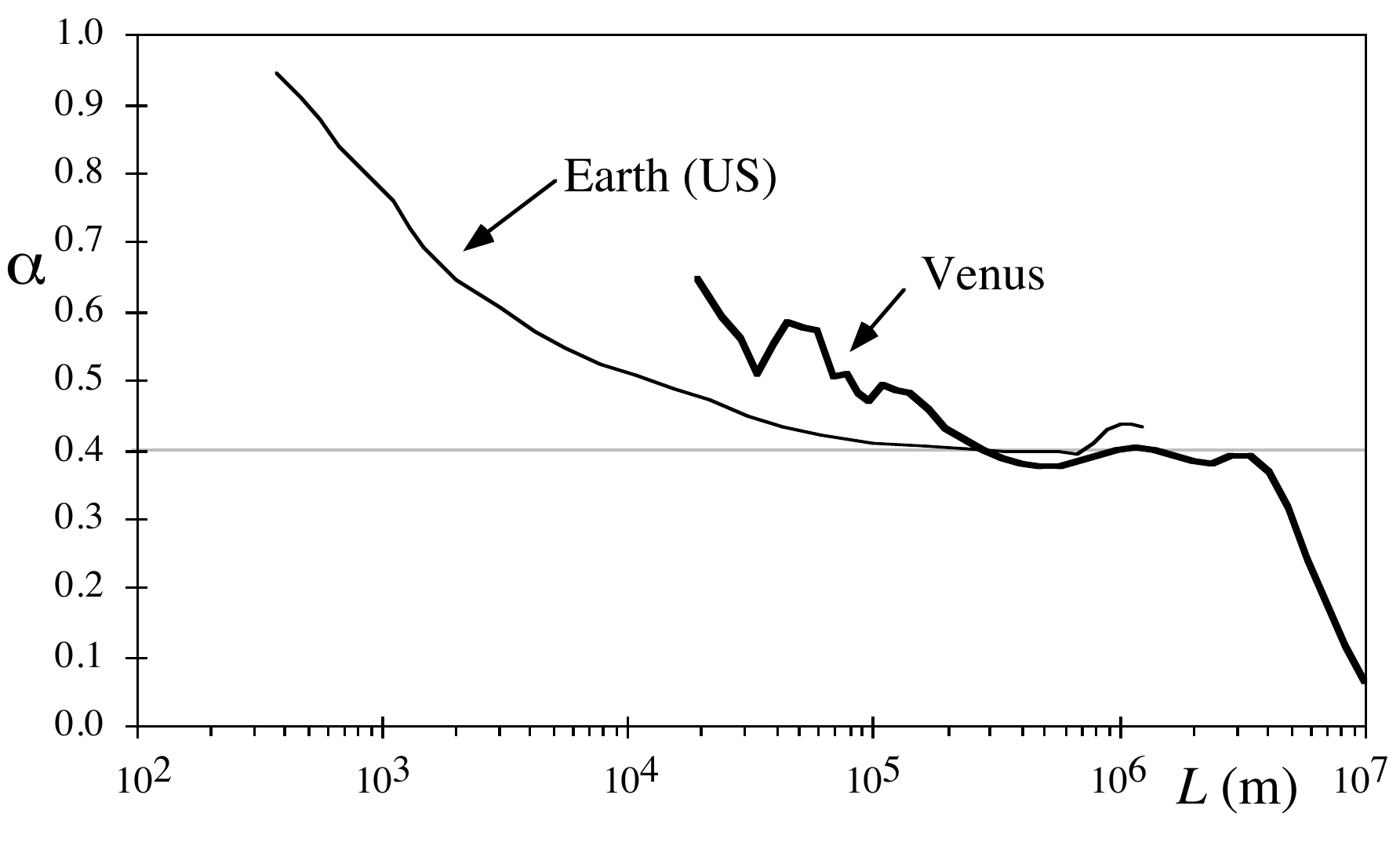}
	\caption{The roughness scaling as a function of length scale for 
	the continental US and Venus data shown in Fig.  2.  The crossover 
	from $\alpha$=1 scaling at small length scales to $\alpha$=0.4 at longer
	 length scales is apparent.  The rapid decrease of $\alpha$ for the Venus
	  data at the longest length scales occurs at scales that approach 
	planetary size.  }
	\label{Fig:3}
\end{figure}

Our results for Venus are also consistent with previous results.  
Turcotte has analyzed the large scale topography using a spherical 
harmonic expansion and obtained exponents\cite{Kucinskas94,Turcotte97} 
of $\alpha$=0.37 and $\alpha$=0.5 
based on two fits of his data, which has significant scatter.  This 
range is consistent with our results.  Turcotte reports that Venus is 
less rough than the Earth when bathymetric and topographic data are 
included together.  Consistent with the previous discussion we find 
the Earth topography, limited to the continental US, to be less rough 
than the Venus topography.

 The observation of crossover behavior in both Earth and Venus leads 
 to questions about the mechanisms that cause the crossover and the 
 origin of the characteristic length scales, which are distinct on 
 Earth (5 km) and Venus (50 km).  This can only be understood once the 
 processes responsible for the two different regimes are understood.  
 The large scale behavior follows the expected universal scaling for 
 up-down asymmetric dynamic surfaces, and thus might be assumed 
 justified even if detailed mechanisms are not well established.  
 However, the short length behavior must still be understood.  It 
 seems natural to attribute the highly correlated smooth behavior of 
 the Earth and Venus at the shortest length scales to erosional 
 processes.  However, recent work\cite{Czirok93,Takayasu92} has 
 suggested that erosion produces self-affine behavior, which would not 
 be consistent with the observed results.  Thus, if erosion is indeed 
 responsible for the short range scaling behavior, the relevant 
 mechanisms of erosion must be better understood.

We would like to thank D. L. Turcotte and M. Kardar for helpful 
comments.

%%%%% References. %%%%%%

%\nocite{*}
%\bibliographystyle{prsty}
%\bibliography{fractLand}

\end{document}